\documentclass[a4paper,fleqn,usenatbib]{mnras}
\usepackage{newtxtext,newtxmath}
\usepackage[T1]{fontenc}
\usepackage{ae,aecompl}
\usepackage{epsfig}
\usepackage{amsmath, amsfonts, epsfig, xspace}
\usepackage{natbib}
\usepackage{amssymb}
\usepackage{rotating}
\usepackage{graphicx}
\usepackage{caption}
\usepackage{longtable}
\usepackage{multicol}
\usepackage{booktabs}
\usepackage{amsmath}
\usepackage{rotating}
\usepackage{changepage}
\def\balblzs{BAL$-$blazars}
\def\balblz{BAL$-$blazar}
\def\apj{{ApJ}}%
\def\apjl{{ApJ}}%
\def\apjs{{ApJS}}%
\def\aap{{A\&A}}%
\def\mnras{{MNRAS}}%
\def\pasp{{PASP}}%
\def\pasj{{PASJ}}%

\title[INOV study of \balblzs]{Are there broad absorption-line blazars?}
\author[$Mishra$ et al.]{{\Large Sapna Mishra$^{1,2}$\thanks{E-mail: sapna@aries.res.in(SM)},
Gopal-Krishna$^{1}$, Hum Chand$^{1}$, Krishan Chand$^{1}$, Vineet Ojha$^{1}$}\\\\
  $^{1}$Aryabhatta Research Institute of Observational Sciences (ARIES), Manora Peak, Nainital $-$ 263002, India\\
  $^{2}$Department of Physics \& Astrophysics, University of Delhi, Delhi 110 007\\}
 
\begin{document}
\date{Accepted ---. Received ---; in original form ---}
\pagerange{\pageref{firstpage}--\pageref{lastpage}} \pubyear{2019}
\maketitle
\label{firstpage}
\begin{abstract}
We report the first systematic search for blazars among
broad-absorption-line (BAL) quasars. This is based on our intranight
optical monitoring of a well-defined sample of 10 candidates selected
on the criteria of a flat spectrum and an abnormally high linear
polarization at radio wavelengths. A small population of BAL blazars
can be expected in the `polar model' of BAL quasars. However, no such
case is found, since none of our 30 monitoring sessions devoted to the
10 candidates yielded a positive detection of intra-night optical
variability (INOV), which is uncharacteristic of blazars. This
  lack of INOV detection contrasts with the high duty cycle of INOV
  observed for a comparison sample of 15 `normal' (i.e., non-BAL)
  blazars. Some possible implications of this are pointed out.
\end{abstract}


\section{Introduction}
\label{introduction}
Blue-shifted broad absorption-line (BAL) troughs seen in the
optical/UV spectra \citep[][]{1991ApJ...373...23W} of $\sim$20\% of
QSOs are interpreted as the covering factor of the BAL outflow in the
orientation based models, or as the duration of BAL phase, in a
QSOs life in the evolutionary models \citep[e.g.,][and references
  therein] {2012A&A...548A..66P}. The outflow speed is found to be as
high as 0.3c \citep[e.g.,][]{2018MNRAS.476..943H}.  In $\sim$
1/7th of BALQSOs, the thermal plasma outflow is accompanied by
ejection of a pair of relativistic plasma jets which can extend to kpc
scale \citep[][]{2000ApJ...538...72B}.  Inferring the orientation of
the jet axis from the spectral/structural radio properties as a
statistical indicator has revealed no preference between the
postulated equitorian
\citep{1995ApJ...448L..77C,1995ApJ...448L..73G,1995ApJ...451..498M,
  2000ApJ...545...63E,2000ApJ...543..686P} and polar
\citep{1999ApJ...527..609P,
  2000ApJ...538...72B,2006ApJ...639..716Z,2007ApJ...661L.139G,2009PASJ...61.1389D}
configurations for the BAL outflows \citep[see, e.g.,][]
{2003A&A...397L..13J,2006MNRAS.372L..58B,2006ApJ...641..210G,
  2013A&A...554A..94B,2015A&A...579A.109K}.  Although, both
configurations may conceivably exist in a single BALQSO
\citep[e.g.,][]{2006MNRAS.372L..58B,2012MNRAS.419L..74Y}, the polar
configuration would have a more direct bearing on the relativistic
jet, given the likelihood of a physical interaction between the
outflowing BAL clouds and the jet on the inner parsec scale. Since the
radio flux of such compact relativistic jets, when pointed near the
line of sight, would appear strongly Doppler boosted, the polar
outflows of absorbers should be frequently observed in those BALQSOs
whose radio spectrum is flat or inverted, and this indicator has been
employed in several studies
\citep{2000ApJ...538...72B,2008MNRAS.388.1853M,bruni12}.  Additional
examples of such BALQSOs with aligned jets have been found through
radio flux variability
\citep{2006ApJ...639..716Z,2007ApJ...661L.139G,2008MNRAS.388.1853M,2009PASJ...61.1389D}.\par

Interestingly, Fanaroff-Riley type II radio morphology is $\sim$ 10
times rarer among BAL quasars, compared to normal quasars in the Sloan
Digital Sky Survey (SDSS) \citep[see ][]{2006ApJ...641..210G}. The
core dominated flat-spectrum subpopulation of radio-loud quasars,
called FSRQs, has a subset, namely blazars, which is characterised by
parsec-scale relativistic jets of strongly Doppler boosted nonthermal
radiation often showing superluminal motion, in addition to rapid flux
variability and a strong (p$_{opt} > 3\%$) optical polarization which
is highly
variable\citep[e.g.,][]{1988A&A...205...86F,2000ApJ...541...66L}.
Another well established exceptionality to blazars is their strong
intra-night optical variability (INOV), of amplitude $\psi >$ 3-4\%
with a duty cycle of around 40-50\% \citep[][and references
  therein]{2018BSRSL..87..281G}. This opens up the possibility to
  confirm the blazar nature of the BAL quasars which are known to
  exhibit at least some radio properties that are commonly associated
  with blazars and hence may be regarded as manifestations of the
  polar model of the BAL phenomenon mentioned above. Here we shall
apply the INOV test to a well-defined sample of 10 radio-loud BALQSOs
exhibiting blazar signatures, namely a flat/inverted radio spectrum
and a large radio polarization that locates them in the high
  polarisation tail for BAL quasars (optical polarimetric data on
  radio-loud BALQSOs being even scarcer, at present). This selection
  process makes our sample particularly suited for making a search for
  `BAL-blazars', in contrast to a previous INOV search which was
  focussed on BAL quasars selected on the criterion of radio loudness
  alone and was found to show only a muted INOV \citep[unlike
    blazars, e.g. see,][] {2013MNRAS.429.1717J}.
\section{The sample of BAL$-$blazar candidates}
\label{sample}
Our sample of 10 \balblz~ candidates for intranight optical monitoring
was derived from 6 publications reporting 56 radio detected BALQSOs
(Table~\ref{sample_blz}).  Out of these, we have selected all 10
sources which have a positive declination, an SDSS-r band apparent
magnitude, m$_{r}$ $<$ 19, a flat or inverted radio spectrum
  \citep[i.e., $\alpha > -$0.5,][]{2000ApJ...538...72B} and a radio
(linear) polarization $p_{rad} > $ 3\% (Table~\ref{sample_blz}).  The
choice of the $p_{rad}$ = 3\% threshold is based on the distribution
of median fractional polarization of the radio core, measured for 387
AGNs, by multi-epoch VLBI at 15 GHz \citep[figure 1
  in][]{2018ApJ...862..151H}.  This distribution consists of a single
large bump peaking at $p_{rad}$(15 GHz) = 1.5\%, followed by a sharp
drop setting in at $p_{rad}$(15 GHz) $\sim$ 3.0\%, and finally
culminating in a low-amplitude tail which extends up to
$p_{rad}$(15GHz) = 9\%. This high polarization tail of FSRQs is
strikingly similar to the distribution of $p_{rad}$(15 GHz) found for
BL Lacs, which is plotted in the same figure 1
\citep[][]{2018ApJ...862..151H} and for which the median value of
$p_{rad}$(15 GHz) = 3.5\%. Hence, it seems reasonable to expect that
at least some of the flat-spectrum BAL quasars falling within the high
polarization tail would turn out to be the putative \balblzs.
\begin{table*}

\caption{\scriptsize The sample of 10 \balblz~ candidates monitored for intranight optical variability (INOV).}
\label{sample_blz}
\begin{adjustwidth}{-1.1cm}{}
\tiny
{
  \begin{tabular}{clllllllcrlll}
    \hline
Source name &  RA (J2000)  &  DEC (J2000)  &  $z_{em}$  &  AI          &  m$_{r}$  &  S$_{1.4 GHz}$  &  S$_{150 MHz}$  &  L$_{150 MHz}$ &  $\alpha_{radio}$  &  Log(R)  &  \multicolumn{1}{l}{Fractional}  & Ref.  \\
           &     &  &           &  &          &           &          &      &      &   & polarisation  &  codes  \\
SDSS        &  hh:mm:ss   &  $^{\circ}$:':''&           &  (kms$^{-1}$) &          & mJy           & mJy          &  erg/s/Hz &                  &          & $\%$ (at GHz)  &    \\

 {(1)}  &  {(2)}  &  {(3)}  &  {(4)}  &  {(5)}  &  {(6)}  &  {(7)}  &  {(8)}  &  {(9)}  &  {(10)}  &  {(11)}  &  {(12)}  &  {(13)}  \\     
\hline\hline
J090552.41$+$025931.5 & 09:05:52.41 & +02:59:31.5 &  1.82 & 130        & 16.91     & 35.6         & 46.80          & 1.07$\times$10$^{34}$                &  0.12             &    1.76        &  7.7  $\pm$3.8  ( 1.4 )    & d,f     \\     
J092824.13$+$444604.7 & 09:28:24.13 & +44:46:04.7 &  1.90 & 293        & 18.31     & 166.5        & 73.60          & 1.89$\times$10$^{34}$                & $-$0.36           &    2.80        &  4.6  $\pm$0.7  ( 1.7 )    & c,e     \\     
J092913.97$+$375743.0 & 09:29:13.97 & +37:57:43.0 &  1.91 & 2170       & 18.04     & 42.7         & 125.30         & 3.24$\times$10$^{34}$                &  0.48             &    2.20        &  6.1  $\pm$0.9  ( 1.4 )    & d,f     \\     
J105416.51$+$512326.1 & 10:54:16.51 & +51:23:26.1 &  2.34 & 2220       & 18.73     & 34.8         & 74.10          & 3.14$\times$10$^{34}$                &  0.34             &    2.38        &  4.2  $\pm$1.1  ( 1.4 )    & d,f     \\     
J115944.82$+$011206.9 & 11:59:44.82 & +01:12:06.9 &  2.00 & 2887       & 17.27     & 271.1        & 331.10         & 9.57$\times$10$^{34}$                &  0.09             &    2.60        &  6.5  $\pm$1.7  ( 1.4 )    & c,b,e   \\     
J123717.44$+$470807.0 & 12:37:17.44 & +47:08:07.0 &  2.27 & 1300       & 18.58     & 89.2         & 50.30          & 1.98$\times$10$^{34}$                & $-$0.26           &    2.47        &  5.9  $\pm$2.6  ( 8.46 )   & d,f     \\     
J131213.57$+$231958.6 & 13:12:13.57 & +23:19:58.6 &  1.51 & -          & 17.32     & 45.8         & 67.90          & 0.98$\times$10$^{34}$                &  0.18             &    1.88        &  3.6  $\pm$2.3  ( 22 )     & b       \\      
J140653.84$+$343337.3 & 14:06:53.84 & +34:33:37.3 &  2.56 & 350        & 18.72     & 165.3        & 150.00         & 7.91$\times$10$^{34}$                & $-$0.04           &    2.83           &  3.5  $\pm$0.2  ( 8.46 )   & d,f     \\     
J162453.47$+$375806.6 & 16:24:53.47 & +37:58:06.6 &  3.38 & 1020       & 18.45     & 54.6         & 35.60          & 3.67$\times$10$^{34}$                & $-$0.19           &    2.41        &  11.3 $\pm$1.5  ( 22.46 )  & a,b,d,f \\     
J162559.90$+$485817.5 & 16:25:59.90 & +48:58:17.5 &  2.72 & -          & 18.09     & 25.8         & $<$7.00        & 0.55$\times$10$^{34}$                & $-$0.47           &    1.90        &  17.6 $\pm$14.0 ( 43 )     & b       \\       
\hline
\multicolumn{13}{l}{\scriptsize Notes. Col. 1: source name; Col . 2,3: coordinates; Col. 4: emission redshift; Col. 5: absorption index \citep{2002ApJS..141..267H}; Col. 6: r-band magnitude from SDSS;}\\
\multicolumn{13}{l}{\scriptsize Col. 7: peak flux density at 1.4 GHz; Col. 8,9: peak flux density and luminosity at 150 MHz; Col. 10: spectral index ($f_{\nu} \propto \nu^{\alpha}$) betweeen 150 MHz and 1.4 GHz;} \\
\multicolumn{13}{l}{\scriptsize Col. 11: radio loudness parameter $R = f_\mathrm{5GHz}/f_\mathrm{2500\text{\normalfont\AA}}$; Col 12. fractional polarization (frequency); Col. 13: reference code(s).}\\
\multicolumn{13}{l}{\scriptsize References: (a) \citet{2005MNRAS.360.1455B}; (b) \citet{2008MNRAS.388.1853M}; (c) \citet{2009PASJ...61.1389D}; (d) \citet{bruni12}; (e) \citet{2013ApJ...772....4H}; (f) \citet{bruni15}.}\\

  \end{tabular}
  }
  \end{adjustwidth}
\end{table*}
\section{Intranight photometric monitoring}
\label{monitoring}
Photometric monitoring of each of our 10 \balblz~ candidates was
performed on 3 nights, adding up to a total of 30 intra-night
monitoring sessions (details in the online tables 1 and 2).  For 25 of
the 30 sessions we used the 1.3-meter Devasthal Fast Optical Telescope
(DFOT), equipped with a Peltier-cooled Andor CCD camera which has 2048
$\times$ 2048 pixels of 13.5 $\mu$m size, providing 18 arcmin
field-of-view (FoV) on the sky \citep[][] {2011CSci..101.1020S}. For
another 3 sessions the 1.04 meter Sampurnanand Telescope (ST) was
used, which is equipped with a 1340 $\times$ 1300 pixel
liquid-nitrogen cooled PyLoN CCD of 20-micron pixel size, providing a
6.8 $\times$ 6.5 arcmin FoV \citep[][]{1999CSci...77..643S}.
Monitoring in the remaining 2 sessions was carried out with the
2.0-meter Himalayan Chandra Telescope (HCT) equipped with a 2148
$\times$ 2048 cryogenically cooled CCD detector covering a 10 $\times$
10 arcmin FoV \citep[][]{2010ASInC...1..193P}.  Each target AGN
(\balblz~ candidate) was monitored in 3 separate sessions continuously
for a minimum duration of 3-hours, in a sequence of 5-10 minute long
exposures. At least 2 or 3 comparison stars were also recorded on each
CCD frame, enabling differential photometry of the target AGN relative
to the comparison stars which were chosen {\it a priori} on the basis
of their proximity to the AGN, both in apparent magnitude and the
location on the CCD chip (online table~1).\par
\begin{figure}
\hspace{-0.9in}
\includegraphics[trim={0 7.5cm 0 0},clip,width=0.7\textwidth,height=0.30\textheight]{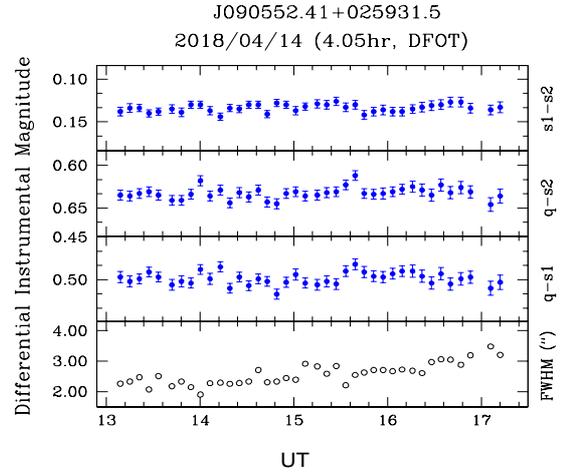}
\caption{An example of the DLCs obtained in the present study. The
  target AGN is the BAL quasar J090552.41$+$025931.5. DLCs for all the
  30 sessions are presented in online figure~1. The date and duration of
  monitoring and the telescope used are mentioned at the top. The
  profile in the upper panel presents the DLC of the chosen two
  comparison stars (``star-star" DLC ).  The two middle profiles
  display the DLCs of the BAL quasar relative to the two comparison
  stars, as mentioned in the labels on the right side.  The profile in
  the bottom panel shows the variation of the seeing disk (FWHM)
  through the monitoring session. }
\label{fig:J0905}
\end{figure}

 Preliminary image
processing (bias subtraction, flat-fielding, and cosmic-ray removal)
was performed using the standard packages in IRAF\footnote{Image
  Reduction and Analysis Facility, distributed by NOAO, operated by
  AURA, Inc. under agreement with the US NSF.}. Aperture photometry
was carried out using the task DAOPHOT~{\sc ii} \citep[Dominion
  Astrophysical Observatory Photometry
  software,][]{1987PASP...99..191S}.  Signal-to-noise ratio was found
to peak for a photometric aperture radius nearly 2 times the seeing
disc whose FWHM was determined by averaging the profiles of 5 fairly
bright, unsaturated stars recorded in the same CCD frame. The measured
instrumental magnitudes were then used to derive the `differential
light curves' (DLCs) of the target AGN relative to two (steady)
comparison stars, as well as the star$-$star DLC for the session
\citep[see][for details of the procedure] {2013MNRAS.435.1300G}.  In
Fig.~\ref{fig:J0905}, we show the differential light curves (DLCs)
obtained from one of the 30 sessions; the DLCs for all 30 monitoring
sessions are presented in the online figure~1.
\section{Statistical analysis and results}
\label{stat_ana}
Following \citet{2013MNRAS.435.1300G}, we applied the widely used
F$^{\eta}$ test to the DLCs of the target AGN relative to the two
comparison stars (s1, s2).  The parameter F$^{\eta}$ computed for the
two AGN DLCs, is defined as:
\begin{equation}
  \label{eq.ftest2}
  \hspace{0.2in} F_{j}^{\eta} = \frac{Var(q-s_{j})}{ \eta^2 \langle \sigma^2_{err} \rangle};  \\
  \langle \sigma^2_{err} \rangle = \sum_\mathbf{i=1}^{N}\sigma^2_{i,err}(q-s_{j})/N
\end{equation}
Here $Var(q-s_{j})$ is the variance of the `AGN$-$j$^{th}$ comparison
star' DLC, where j=1,2 and $\sigma_{i,err}(q-s_{j})$ are the photometric error
of the N individual data points in the DLC (as returned by the DAOPHOT
routine). For each of the two DLCs, i=1 to N and $\eta$ $=$
1.5\citep[see][]{2012A&A...544A..37G}.\par
For each of the two DLCs of the target AGN, online table~2 (Column 7)
compares the computed values of F$^{\eta}$ with the critical value of
F ( = $F_{c}^{\alpha}$). The values of $\alpha$ are set at 0.05 and
0.01, corresponding to 95\% and 99\% confidence levels for INOV
detection.  If the computed F$^{\eta}$ for a DLC exceeds
$F_{c}^{\alpha}$, the null hypothesis (i.e., no variability) is
rejected at the corresponding confidence level. We thus classify a DLC
as variable (`V') if its computed F$^{\eta}$ $>$ F$_{c}$(0.99);
probably variable (`PV') if the F$^{\eta}$ lies between F$_{c}$(0.95)
and F$_{c}$(0.99), and non-variable (`NV') in case F$^{\eta}$ $<$
F$_{c}$(0.95) \citep{2013MNRAS.435.1300G}. Column 10 in the
  online table~2 lists the session averaged photometric accuracy of the measured
  differential magnitudes {$\sqrt {\eta^{2} \langle \sigma^2_{err}\rangle}$
  }. It is estimated using the two AGN-star DLCs and is nearly always
better than 3\% (median = 2.0\%).
\section{Discussion and Conclusion}
\label{discussion}
The main result from this study (online table~2) is the non-detection
of INOV in any of the 30 sessions devoted to the present
representative sample of 10 \balblz~ candidates, selected on the basis
of a flat/inverted spectrum and high linear polarization at
centimeter/decimeter wavelengths (Table~\ref{sample_blz}). We shall
now compare this result with the INOV results reported in
\citet{2013MNRAS.435.1300G} for a sample of 24 `normal' (i.e.,
non-BAL) blazars that were monitored in 85 sessions, also in red
filter and subjected to the same F$^{\eta}$ test.  For this, we first
need to extract an appropriate comparison sample out of those 85
sessions, in order to minimise the difference in sensitivity achieved
for that and the present sample of DLCs. The monitoring for the
present sample has been almost entirely performed using the 1.3-m DFOT
(Sect.~\ref{monitoring}), whereas the 24 normal blazars were mostly
monitored with the 1.04-meter ST.\par
Considering this, we adopt the premise that a proper comparison sample
of DLCs of the normal blazars should be formed by matching in terms of
the rms error found for the (non-varying) ``star-star'' DLC for a
given session. Thus, for the present sample of 30 DLCs of the 10
\balblz~ candidates, we have built a comparison sample of 28 DLCs
pertaining to 15 normal blazars, out of the
\citet{2013MNRAS.435.1300G} sample, adopting a tolerance of $\pm
0.5\%$ in the rms error mentioned above.\par Using the data provided
in table 1 of \citet{2013MNRAS.435.1300G} we have computed the INOV
duty cycle (DC) for the sample of `n' DLCs of normal blazars, as:
\begin{equation}
\hspace{0.8in}    
DC = 100\frac{\sum_\mathbf{i=1}^\mathbf{n}
  K^i(1/T_{int}^i)}{\sum_\mathbf{i=1}^\mathbf{n}(1/T_{int}^i)} {\rm
  per cent}
\label{eq:dc} 
\end{equation}
where T$_{int}^i = T_{obs}^{i}(1+z_{em})^{-1}$ is the intrinsic rest
frame monitoring duration corrected for the cosmological redshift,
z$_{em}$.  $K^{i}$ was taken as 1 for a positive detection of INOV in
the $i^{th}$ session, otherwise, it was set to zero. We thus find, the
INOV DC to be 41.2\% for the comparison sample (28 sessions devoted to
15 normal blazars). This high value is in striking contrast to the
non-detection of INOV in any of the 30 sessions devoted to our sample
of 10 \balblz~ candidates.\par

\begin{figure}
  \hspace{-0.3in}
  \includegraphics[width=0.5\textwidth,height=0.38\textheight]{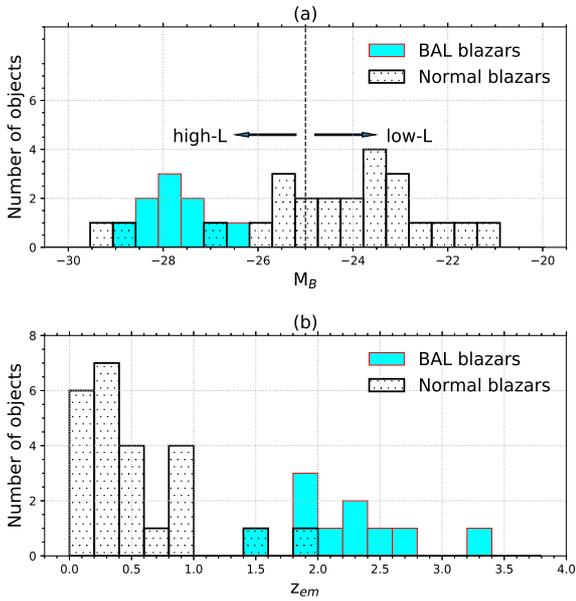}
  \caption{ Distribution of absolute magnitude in B-band (M$_{B}$, upper
    panel) and emission redshift (lower panel) for the present sample
    of 10 \balblz~ candidates (cyan filled) and for the sample of 24
    normal blazars from \citet{2013MNRAS.435.1300G} (black dotted). The dotted vertical in the upper panel
  corresponds to M$_{B}$ = $-$25.0}
  \label{fig:absMag_zem} 
\end{figure}

\begin{figure}
  \vspace{-0.1in}
  \hspace{-0.3in}
\includegraphics[width=0.55\textwidth,height=0.35\textheight]{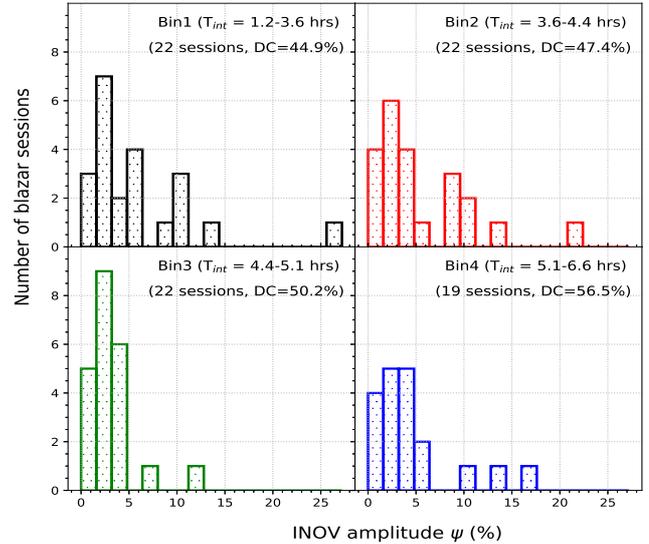}
\caption{Distribution of INOV amplitude ($\psi$) for the 4 bins of T$_{int}$, 
derived using the comparison sample of 24 normal blazars monitored in 85 sessions
(see text).}
\label{fig:INOV_amp} 
\end{figure}

We now examine the possibility that the strong contrast between the
INOV duty cycles found here between the samples of \balblz~ candidates
and normal blazars might arise from the systematic difference between
their optical luminosities and/or redshifts
(Fig.~\ref{fig:absMag_zem}(a), Fig.~\ref{fig:absMag_zem}(b)).  To
check the possible role of luminosity, we turn to the
\citet{2013MNRAS.435.1300G} sample of 85 DLCs of 24 normal blazars and
divide it into two luminosity bins separated in absolute magnitude in
B-band (M$_{B}$) at M$_{B}$ = $-$25.0.  The higher luminosity (high-L)
bin contains 8 normal blazars monitored in 40 sessions and the lower
luminosity (low-L) bin contains 16 normal blazars (45 sessions).  For
a proper comparison between these two sets of sessions, we again apply
the afore-mentioned filter of rms matching to within a tolerance of
$\pm 0.5\%.$ This led to 37 sessions (8 blazars) from the high-L bin
matching in the rms error with 37 sessions (16 blazars) of the low-L
bin.  The corresponding median M$_{B}$ for these two luminosity bins
are $-$25.4 and $-$23.7, respectively. Following Eq.~\ref{eq:dc} we
have computed the INOV duty cycles and find them to be 44.5\% and
52.0\% for the high and low luminosity bins, respectively. Based on
this, any decrement of DC with luminosity appears to be marginal.
Could the marked contrast between the INOV of the \balblz~ candidates
and normal blazars then be because of the redshifts of the \balblz~
candidates being systematically higher than those of the comparison
sample of normal blazars described above (median redshifts of the two
samples are 2.13 and 0.42)? An important consequence of this
difference is that, when translated to the rest-frame, the intranight
monitoring durations (T$_{int}$) are substantially shorter for the
sample of \balblz~ candidates. To seek a clue on this point, we again
turn to the afore-mentioned \citet{2013MNRAS.435.1300G} sample of 85
intranight monitoring sessions devoted to 24 normal blazars. We divide
this dataset into 4 bins of T$_{int}$, each containing close to 20
sessions. Median values of T$_{int}$ for these bins are 3.0, 4.0, 4.8
and 5.6 hr (Fig.~\ref{fig:INOV_amp}). Computation of INOV DC for these
4 bins (Eq.~\ref{eq:dc}) gives INOV DCs of 44.9\%, 47.4\%, 50.2\%, and
56.5\%, respectively (note that each of these estimates is the average
of the DCs calculated for the two DLCs of a given AGN, derived
relative to the two comparison stars).  It is seen that even for the
shortest bin of T$_{int}$ (median = 3.0 hr) the INOV duty cycle is
44.9\%, which is only marginally lower than the DCs found for the 3
bins of longer duration. Furthermore, there is little evidence that
strong INOV is a rarer occurrence in sessions of shorter intrinsic
duration (at least over this range of T$_{int}$). From
Fig.~\ref{fig:INOV_amp} it is seen that for the shortest bin (with
median = 3.0 hr), strong INOV with amplitude $\psi \gtrsim$ 10\% was
detected in 5 out of 22 sessions, which is evidently not smaller than
the corresponding value (7/63) found for the aggregate of the 3 bins
of longer sessions.\par
Thus, in order to reconcile the foregoing analysis with the present
non-detection of INOV for the \balblz~ candidates in any of the 30
sessions (median T$_{int}$ = 1.2 hr) one might postulate a drastic
drop in INOV strength for T$_{int}$ $\lesssim$ 1$-$2 hrs. Although,
such a hypothesis cannot be ruled out at present \citep[see,
  e.g.,][]{2002A&A...390..431R,2011MNRAS.416..101G,2018BSRSL..87..281G},
and it remains observationally verifiable, the alternative possibility
that the non-detection of INOV for the \balblz~ candidates could
either undermine the polar model for the BAL outflows
(Sect.~\ref{introduction}), or be traceable to some other physical
effect associated with that model.  For instance, could it be that, as
compared to normal blazars, physical conditions in the relativistic
jets of BAL quasars are less conducive for strong INOV \citep[as, for
  example, is their propensity to be incapable of developing FR II
  radio structures, e.g. see,][]{2006ApJ...641..210G}.  Origin of INOV has been
associated with a zone of turbulence within a parsec-scale jet, just
upstream of a relativistic shock, as sketched in
\citet{2008Natur.452..966M} \citep[see,
  also,][]{2012A&A...544A..37G,2016ApJ...820...12P}. Conceivably, the
shock induced turbulence in the jets of BAL quasars is not strong
enough to accelerate relativistic particles to the high energies
needed for the emission of optical synchrotron radiation.  Such a
scenario would call for a more refined understanding of the physics of
interaction of the inner relativistic jet with the rapidly outflowing
BAL clouds of much denser thermal plasma, as envisioned in the polar
model of BAL quasars \citep[e.g.,][]{2007ApJ...661L.139G}. In parallel
to such theoretical studies, it would also be worthwhile to intensify
intranight optical monitoring of \balblz~ candidates in sessions of
longer durations.
\section*{Acknowledgments}
We thank the anonymous referee for the constructive comments on our
manuscript. The scientific and technical staff of ARIES DFOT and ST
are deeply acknowledged. Thanks are also due to the staff of IAO
(Hanle) and CREST (Hosakote), for making possible a part of the
observations reported here. The facilities at IAO and CREST are
operated by the Indian Institute of Astrophysics, Bangalore.  The
valuable observing support provided by Ms. Raya Dastidar,
Ms. Mridweeka Singh and Mr. Rakesh Pandey is also highly acknowledged.

\bibliography{references}
\label{lastpage}
\end{document}